# Evaluation of Proactive, Reactive and Hybrid Ad hoc Routing Protocol for various Battery models in VANET using Qualnet

**Manish Sharma[1] & Gurpadam Singh[2]**

[1]Department of Physics, Govt. College, Dhaliara, H.P. India
[2]Deparment of ECE, B.C.E.T., Gurdaspur, Punjab, India
E-mail : manikambli@rediffmail.com[1], gurpadam@yahoo.com[2]

*Abstract -* In VANET high speed is the real characteristics which leads frequent breakdown, interference etc. In this paper we studied various Ad hoc routing protocols, Reactive, Proactive & Hybrid, taking in to consideration various VANET parameters like speed, altitude etc in real traffic scenario and evaluated them for various battery models for energy conservation.. The AODV and DYMO (Reactive), OLSR (Proactive) and ZRP (hybrid) protocols are compared for battery models Duracell AA(MX-1500),Duracell AAA(MN-2400),Duracell AAA(MX-2400), Duracell C-MN(MN-1400),Panasonic AA standard using Qualnet as a Simulation tool. Since Energy conservation is main focus now days. Hence performance of the protocols with various battery models counts and helps to make a right selection. Varying parameters of VANET shows that in the real traffic scenarios proactive protocol performs more efficiently for energy conservation.

## I. INTRODUCTION

Vehicular Ad hoc Network (VANET) is a new communication paradigm that enables the communication between vehicles moving at high speeds. It has been found that mobile terminals in fast moving vehicles like cars, buses, trains are frequent signal breakdowns as compared to pedestrians. It has been found that in the last decade so many functions like gaming, internet etc. has been added leading to fast CPU clock speed hence more battery consumption. In order to improve QoS and energy conservation in fast moving vehicles various light weight routing protocols needed to be studied in Physical and data link layer. So that Right selection of the protocol can be made. There are mainly three types of routing protocols, Reactive [1], Proactive [2], Hybrid [3]. These protocols are having different criteria for designing and classifying routing protocols for wireless ad hoc network. The Mobile Ad hoc Network (MANET) working group of the Internet Engineering Task Force (IETF) [4] develops standards for routing in dynamic networks of both mobile and static nodes. The protocols in focus now days are Hybrid protocols and others [7]. Its use in the context of VANET's along with reactive and proactive has always been area under investigation. Routing protocols are always challenging in the fast moving nodes as their performance degrades and such type of network is difficult to manage as fast handoff, signal quality, Interference maximizes along with other geographical factors.

In this work, the feasibility, the performance, and the limits of ad hoc communication using the three types of protocols is evaluated as per battery models Duracell AA(MX-1500),Duracell AAA(MN-2400),Duracell AAA(MX-2400),Duracell C-MN(MN-1400)[11],Panasonic AA[12] and Potentials for optimizing the deployed transport and routing Protocols is investigated. Special care is taken in to provide Realistic scenarios of both road traffic and network usage. This is accomplished by simulating a scenario with the help of simulation tool Qualnet [6].A micro simulation environment for road traffic supplied vehicle movement information, which was then fed in to an event-driven network simulation that configured and managed a VANET model based on this mobility data. The protocols and their various parameters of the transport, network, data link, and physical layers were provided by well-tested implementations for the networks simulation tool, while VANET mobility is performed by our own implementation.

## II. AD HOC ROUTING PROTOCOLS

Routing protocol is a standard that controls how nodes decide how to route the incoming packets between devices in a wireless domain & further Distinguished in many types. There are mainly three types of routing protocols. Ad-hoc on demand vector distance vector (AODV), Dynamic MANET On demand (DYMO) and Dynamic source routing (DSR) are the examples of reactive routing protocols whereas Optimized Link State Routing (OLSR) and Fisheye state routing (FSR) are the examples of proactive routing protocols. Hybrid routing protocols is the combination of both proactive and reactive routing protocols, Temporary Ordered Routing Algorithm (TORA), Zone Routing Protocol (ZRP), Hazy Sighted Link State (HSLS) and Orderone Routing Protocol (OOPR) are its examples. In our work the chosen protocols are AODV, DYMO, OLSR and ZRP.

### A. Ad-hoc on demand vector distance vector (AODV)

AODV [7] shares DSR's on-demand characteristics in that it also discovers routes on an as needed basis via a similar route discovery process. However, AODV adopts a very different mechanism to maintain routing information. It uses traditional routing tables, one entry per destination. This is in contrast to DSR, which can maintain multiple route cache entries for each destination. Without source routing, AODV relies on routing table entries to propagate an RREP back to the source and, subsequently, to route data packets to the destination. AODV uses sequence numbers maintained at each destination to determine freshness of routing information and to prevent routing loops. All routing packets carry these sequence numbers.

### B. Dynamic MANET On demand (DYMO)

DYMO[9] is another reactive routing protocol that works in multi hop wireless networks. It is currently being developed in the scope of IETF's [4] MANET working group and is expected to reach RFC status in the near future. DYMO is considered as a successor to the AODV routing protocols. DYMO has a simple design and is easy to implement. The basic operations of DYMO protocol are route discovery and route Maintenance was studied extensively [8] along with comparison of two on demand routing protocols.

### C. Optimized Link State Routing (OLSR)

OLSR [10] is the proactive routing protocol that is evaluated in this synopsis. Basically OLSR is an optimization of the classical link state algorithm adapted for the use in wireless ad hoc networks. In OLSR, three levels of optimization are achieved. First, few nodes are selected as Multipoint Relays (MPRs) to broadcast the messages during the flooding process. This is in contrast to what is done in classical flooding mechanism, where every node broadcasts the messages and generates too much overhead traffic.

### D. Zone Routing Protocol (ZRP)

Hybrid routing combines characteristics of both reactive and proactive routing protocols to make routing more scalable and efficient [11]. Mostly hybrid routing protocols are zone based; it means the number of nodes is divided into different zones to make route discovery and maintenance more reliable for MANET. The need of these protocols arises with the deficiencies of proactive and reactive routing and there is demand of such protocol that can resolve on demand route discovery with a limited number of route searches. ZRP limits the range of proactive routing methods to neighbouring nodes locally; however ZRP uses reactive routing to search the desired nodes by querying the selective network nodes globally instead of sending the query to all the nodes in network. ZRP uses "Intrazone" and "Interzone" routing to provide flexible route discovery and route maintenance in the multiple ad hoc environments. Interzone routing performs route discovery through reactive routing protocol globally while intrazone routing based on proactive routing in order to maintain up-to-date route information locally within its own routing range . The overall characteristic of ZRP is that it reduces the network overhead that is caused by proactive routing and it also handles the network delay that is caused by reactive routing protocols and perform route discovery more efficiently. Normal routing protocols which works well in fixed networks does not show same performance in mobile ad hoc networks. In these networks routing protocols should be more dynamic so that they quickly respond to topological changes. There is a lot of work done on evaluating performance of various MANET routing protocols for constant bit rate traffic

## III. BATTERY MODELS

The zinc/potassium hydroxide/manganese dioxide cells, commonly called alkaline[12] or alkaline-manganese dioxide cells, have a higher energy output than zinc-carbon (Leclanche) cells. Other significant advantages are longer shelf life, better leakage resistance, and superior low temperature performance. In comparison to the zinc-carbon cell, the alkaline cell delivers up to ten times the ampere-hour capacity at high and continuous drain conditions, with its performance at low temperatures also being superior to other conventional aqueous electrolyte primary cells. Its more effective, secure seal provides excellent resistance to leakage and corrosion.

The use of an alkaline electrolyte, electrolytic ally prepared manganese dioxide, and a more reactive zinc powder contributes to a higher initial cost than zinc-carbon cells. However, due to the longer service life, the alkaline cell is actually more cost-effective based upon cost-per-hour usage, particularly with high drains and continuous discharge. The high-grade, energy-rich materials composing the anode and cathode, in conjunction with the more conductive alkaline electrolyte, produce more energy than could be stored in standard zinc carbon cell sizes In comparison to the zinc-carbon cell, the alkaline cell [13] delivers up to ten times the ampere-hour capacity at high and continuous drain conditions, with its performance at low temperatures also being superior to other conventional aqueous electrolyte primary cells. Its more effective, secure seal provides excellent resistance to leakage and corrosion. The product information and test data included in this section represent Duracell's newest alkaline battery products.

### A. Duracell AA (MX-1500)

| Nominal Voltage: | 1.5  V |
|---|---|
| Operating Voltage | 1.6 - 0.75V |
| Impedance: | 81  m-ohm @ 1kHz |
| Typical Weight: | 24 gm (0.8 oz.) |
| Typical Volume: | 8.4 cm 3  (0.5 in.3) |
| Storage Temperature Range | $-20^{o}$C to $35^{o}$C |
| Operating Temperature Range: Terminals: | $-20^{o}$C to $54^{o}$C  Flat |
| ANSI: IEC: | 15A LR6 |

### B. Duracell AAA (MX-2400)

| Nominal Voltage: | 1.5  V |
|---|---|
| Operating Voltage | 1.6 - 0.75V |
| Impedance: | 114  m-ohm @ 1kHz |
| Typical Weight: | 11 gm (0.4 oz.) |
| Typical Volume | 3.5 cm 3  (0.2 in.3) |
| Storage Temperature Range | $-20^{o}$C to $35^{o}$C |
| Operating Temperature Range: Terminals: | $-20^{o}$C to $54^{o}$C Flat |
| ANSI: IEC: | 24A LR03 |

### C. Duracell AAA (MN-2400)

| Nominal Voltage: | 1.5  V |
|---|---|
| Operating Voltage | 1.6 - 0.75V |
| Impedance: | 250  m-ohm @ 1kHz |

| Typical Weight: | 11 gm (0.4 oz.) |
|---|---|
| Typical Volume: | 3.5 cm 3  (0.2 in.3) |
| Storage Temperature Range | $-20^{o}$C to $35^{o}$C |
| Operating Temperature Range: Terminals: | $-20^{o}$C to $54^{o}$C Flat |
| ANSI: IEC: | 24A LR03 |

### D. Duracell C-MN (MN-1400)

| Nominal Voltage: | 1.5  V |
|---|---|
| Operating Voltage | 1.6 - 0.75V |
| Impedance: | 136  m-ohm @ 1kHz |
| Typical Weight: | 139 gm (4.9 oz.) |
| Typical Volume: | 3.5 cm 3  (0.2 in.3) |
| Storage Temperature Range | $-20^{o}$C to $35^{o}$C |
| Operating Temperature Range: Terminals: | $-20^{o}$C to $54^{o}$C Flat |
| ANSI: IEC: | 13A LR20 |

### E. Panasonic AA

| Nominal Voltage: | 1.5  V |
|---|---|
| Operating Voltage | 1.6 - 0.75V |
| Impedance: | 136  m-ohm @ 1kHz |
| Typical Weight: | 0.80gm (23.0oz.) |
| Typical Volume: | 3.8 cm3 (0.2 in.3) |
| Storage Temperature Range | $-20^{o}$C to $35^{o}$C |
| Operating Temperature Range: Terminals: | $-20^{o}$C to $54^{o}$C Flat |
| ANSI: IEC: | 24A LR03 |

## IV.  SIMULATION TOOL

The collaboration of imminent research objectives and its related scope in this study are also collapsed into same influence of simulation environment for generating some authenticated outcomes. For this purpose, the adopted methodology for the results of this research work (specifically comparative routing analyses) is based on simulations near to the real time packages before any actual implementation.

QualNet is a comprehensive suite of tools for modelling large wired and wireless networks. It uses simulation and emulation to predict the behaviour and performance of networks to improve their design, operation and management. QualNet enables users to Design new protocol models, Optimize new and existing models, Design large wired and wireless networks using pre-configured or user-designed models, Analyze the performance of networks and perform what-if analysis to optimize them. QualNet (6) is the preferable simulator for ease of operation. So, we found QualNet be the best choice to implement our scenarios as we do not need every feature possible, just those for the token passing and message routing. QualNet is a commercial simulator that grew out of GloMoSim, which was developed at the University of California, Los Angeles, UCLA, and is distributed by Scalable Network Technologies [6]. The QualNet simulator is C++ based. All protocols are implemented in a series of C++ files and are called by the simulation kernel. QualNet comes with a java based graphical user interface (GUI).

Table 1. Simulation Parameters

| Simulator | Qualnet Version 5.o.1 |
|---|---|
| Terrain Size | 1500 x 1500 |
| Simulation time | 3000s |
| No. Of Nodes | 15 |
| Mobility | Random Way Point Pause time= 0s |
| Speed of Vehicles | Min.=3m/s Max.=20m/s |
| Routing Protocols | AODV,DYMO,OLSR,ZRP |
| Medium Access protocol | 802.11 MAC, 802.11 DCF Tx Power=150dbm |
| Data size | 512 bytes |
| Data Interval | 250ms |
| No. of sessions | 5 |
| Altitude | 1500 |
| Weather mobility | 100ms |
| Battery models | Duracell AA(MX-1500),Duracell AAA(MN-2400), Duracell AAA(MX-2400), Duracell C-MN(MN-1400), Panasonic AA |

## V. DESIGNING OF SCENARIO

The scenario is designed in such a way that it undertakes the real traffic conditions. We have chosen 15 fast moving vehicles in the region of 1500X1500 with the random way point mobility model. There is also well defined path for some of the vehicles, so that real traffic conditions can also be taken care of. It also shows wireless node connectivity of few vehicles using CBR application. The area for simulation is Hilly area with altitude of 1500 meters. Weather mobility intervals is 100ms.Pathloss model is two ray with max prop distance of 100m.

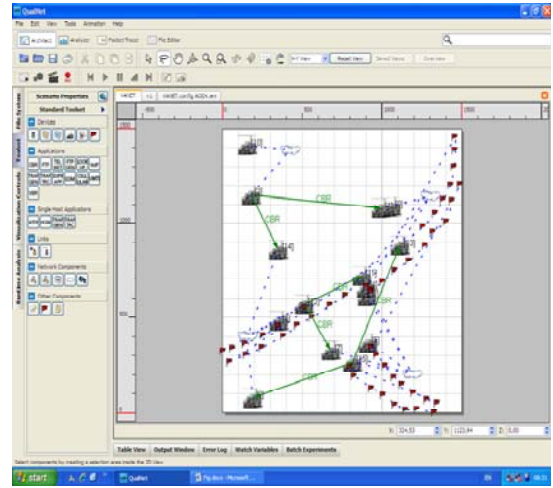

Fig. 1 : Qualnet VANET Scenario

## VI. RESULTS AND DISCUSSION

The simulation result brings out some important characteristic differences between the routing protocols. In all the simulation results OLSR outperforms the other protocols. This is because OLSR is a proactive protocol and it pre determines the route in well defined manner. It uses destination sequence numbers to ensure loop freedom at all times and it offers quick convergence when the network topology changes. The residual battery capacity of OLSR for all the Duracell models are maximum and same whereas for Panasonic model is low as compare for AODV.

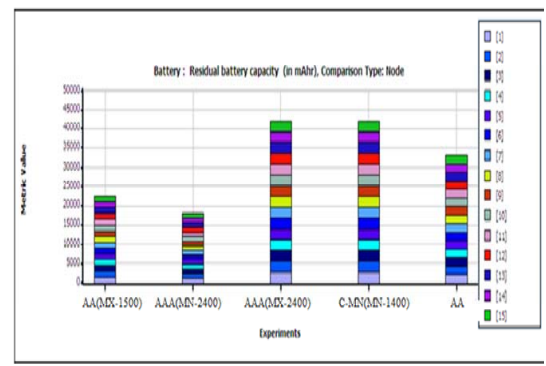

Fig. 2 : Battery model comparison for AODV

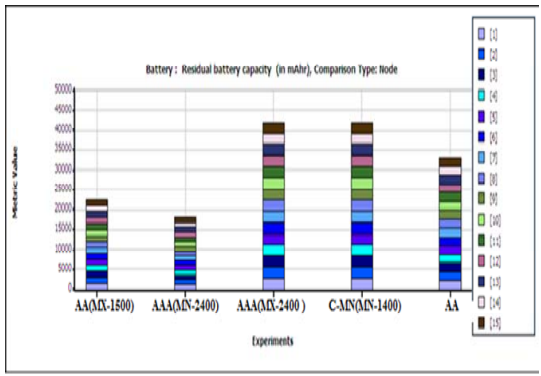

Fig. 3 : Battery model comparison for DYMO

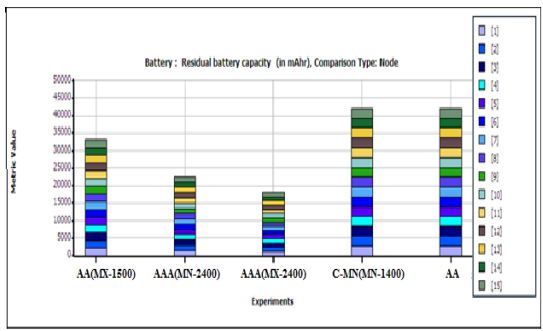

Fig. 4 : Battery model comparison for ZRP

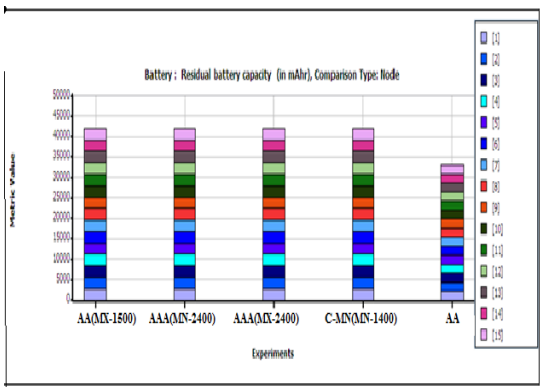

Fig. 5 : Battery model comparison for OLSR

## ACKNOWLEDGEMENT

We are thankful to Nidhi Sharma lecturer HPES for her support and guidance.